\documentclass[runningheads]{llncs}
\usepackage[T1]{fontenc}
\usepackage{amsmath}
\usepackage{multirow}
\usepackage{subcaption}
\usepackage{amsfonts}

\usepackage{graphicx}
\begin{document}
\title{Enhancing Alzheimer's Diagnosis: Leveraging Anatomical Landmarks in Graph Convolutional Neural Networks on Tetrahedral Meshes}
\titlerunning{Landmark-Enhanced Tetrahedral Convolutional Neural Networks}
%
\author{Yanxi Chen\inst{1,\textsuperscript{\textdagger}} \and Mohammad Farazi\inst{1,\textsuperscript{\textdagger}} \and Zhangsihao Yang\inst{1} \and Yonghui Fan\inst{2} \and Nicholas Ashton\inst{3} \and Eric M Reiman\inst{3} \and Yi Su\inst{3} \and Yalin Wang\inst{1,*}}


%
\authorrunning{Y. Chen et al.}
%
\institute{School of Computing and Augmented Intelligence, Arizona State University, Tempe, AZ, USA \and Amazon AGI, Redmond, WA, USA \and Banner Alzheimer’s Institute, Phoenix, AZ, USA}
\maketitle              
\def\thefootnote{\textsuperscript{\textdagger}}\footnotetext{These authors contributed equally to this work}
\def\thefootnote{*}\footnotetext{Corresponding author(s)}

\begin{abstract}
Alzheimer's disease (AD) is a major neurodegenerative condition that affects millions around the world. As one of the main biomarkers in the AD diagnosis procedure, brain amyloid positivity is typically identified by positron emission tomography (PET), which is costly and invasive. Brain structural magnetic resonance imaging (sMRI) may provide a safer and more convenient solution for the AD diagnosis. Recent advances in geometric deep learning have facilitated sMRI analysis and early diagnosis of AD. However, determining AD pathology, such as brain amyloid deposition, in preclinical stage remains challenging, as less significant morphological changes can be observed. As a result, few AD classification models are generalizable to the brain amyloid positivity classification task. Blood-based biomarkers (BBBMs), on the other hand, have recently achieved remarkable success in predicting brain amyloid positivity and identifying individuals with high risk of being brain amyloid positive. However, individuals in medium risk group still require gold standard tests such as Amyloid PET for further evaluation. Inspired by the recent success of transformer architectures, we propose a geometric deep learning model based on transformer that is both scalable and robust to variations in input volumetric mesh size. Our work introduced a novel tokenization scheme for tetrahedral meshes, incorporating anatomical landmarks generated by a pre-trained Gaussian process model. Our model achieved superior classification performance in AD classification task. In addition, we showed that the model was also generalizable to the brain amyloid positivity prediction with individuals in the medium risk class, where BM alone cannot achieve a clear classification. Our work may enrich geometric deep learning research and improve AD diagnosis accuracy without using expensive and invasive PET scans.

\end{abstract}
\section{Introduction}

Alzheimer’s disease (AD) has been recognized as a major health problem since the past century, affecting patient cognition and causing dementia~\cite{qiu2009epidemiology,el2019tip}. Although there is still no cure for AD, several treatments have been developed to alleviate clinical symptoms and slow the disease progression, necessitating the development of early AD diagnosis techniques~\cite{atri2019current,liss2021practical}. Among brain imaging techniques, magnetic resonance imaging (MRI) has emerged as a leading and cost-effective method for detecting brain structural changes associated with AD~\cite{dickerson2009cortical}. Compared to more specialized techniques like positron emission tomography (PET), MRI is less invasive, less costly, and more accessible~\cite{liss2021practical,atri2019alzheimer}.

While MRI images can be represented by 3-D voxels, the limited grid resolution prevents detailed study of arbitrarily small regions~\cite{fawaz2021benchmarking}. Volumetric mesh data structures, such as tetrahedral meshes, offer comprehensive insights by modeling both surface and interior aspects of substances, and have been widely used in scientific modeling such as finite element analysis and physical simulations~\cite{chentanez2007liquid}. Graph neural networks (GNNs) present a natural choice when analyzing graphs, which lack irregular grid-like structures as images. While algorithms based on graph convolution have developed for more commonly used surface meshes~\cite{hanocka2019meshcnn}, convolutions on surface meshes suffer from the over-squashing nature of the message-passing neural networks, potentially aggregating unnecessary or redundant information from intermediate nodes when interacting with long-range nodes. This drawback can naturally be resolved by using volumetric mesh representation, as more detours are available in pathfinding. One notable method for tetrahedral mesh analysis is TetCNN, which was proposed as an improved graph neural networks for tetrahedral mesh data analysis~\cite{farazi2023tetcnn} by replacing graph Laplacian with volumetric Laplace Beltrami operator (volumetric LBO)~\cite{wang2004volumetric,wang2017towards}.

Attention mechanism and transformer architectures are being used ubiquitously in deep learning applications~\cite{vaswani2017attention,dosovitskiy2020image,velivckovic2017graph}. Fundamental to the attention mechanism is a set of vectors associated with the input data, namely query vectors, key vectors, and value vectors. The outputs of each attention layer consist of the weighted sum of the value vectors, where the weights are determined by attention scores representing the similarity between the query and key vectors. Transformer architectures have proved their superiority over traditional CNNs in various tasks, such as machine translation, text generation, image classification and graph classification. However, these methods are not directly suitable for tetrahedral meshes with varying sizes and topologies, necessitating a novel transformer-based framework tailored to this structure with strong scalability.


To address the challenge of applying a transformer to tetrahedral meshes with raw 3D coordinate inputs, we propose generating tokens based on learned local features. Unlike ViT for images, generating patches in tetrahedral meshes is not straightforward. To overcome this, we identify key landmarks (super nodes) using a pre-trained Gaussian Process (GP) model. These landmarks serve as the centers of patches, and an efficient methodology assigns nearest neighbors to each super node, ultimately forming the final patch structure for tokenization. The more important reason we create these patches is to learn features and use a pooling strategy to represent the final token feature, unlike the ViT style models concatenating the raw input without any learning strategy beforehand. This is essential to make sure all patches have the same input dimension.

Recent advancements in blood-based biomarkers (BBBMs), such as plasma Amyloid-$\beta$ 42/40 ratios and phosphorylated tau (pTau) isoforms, offer a promising alternative for detecting AD pathology~\cite{nakamura2018high,ashton2023plasma}. Among these, plasma pTau-217 stands out as the most effective biomarker for classifying brain amyloid positivity and distinguishing AD from other neurodegenerative diseases~\cite{therriault2023equivalence,groot2022diagnostic,palmqvist2020discriminative}. By integrating MRI with BBBMs as auxiliary inputs into our transformer model, we aim to enhance diagnostic performance, particularly for cases where structural imaging alone may be insufficient.

In this study, we proposed landmark-enhanced TetCNN (LETetCNN) for AD diagnosis and brain amyloid positivity classification~\cite{fan2020convolutional}.
\noindent Our primary contributions are summarized as follows:
\begin{enumerate}
    \item We propose, for the first time, a novel tetrahedral mesh autoionization and transformer-based model designed to handle large and varying-sized meshes.
    \item To address the computational challenges of full self-attention, we adopt a sparse local attention strategy, significantly reducing computational cost.
    \item A node-level feature learning module is introduced to improve performance and local contextual awareness before feeding geometry into the transformer.
    \item Our model is rigorously evaluated on subsets of the ADNI dataset for multiple tasks, consistently achieving improvements over state-of-the-art methods.
    \item Integrating BBBMs with learned features significantly boosts performance, highlighting the complementary role of sMRI analysis in the AD studies.
\end{enumerate}

\section{Methods}
\subsection{Notation}
In this paper, we represent the tetrahedral mesh as $M = (\mathit{V},\mathit{T})$, in which $V$ and $T$ represent vertices and tetrahedra set of the volumetric mesh equipped with its corresponding Laplace Beltrami Operator (LBO) $L$. The mesh consists of $N = |V|$ nodes and $E = |E|$ edges and $|P|$ distinct patches after partitioning the mesh. Node-level attributes are denoted by their associated patch numbers as subscripts and is small characters. Accordingly, patch level attributes are stated with capital letters. Layer-level attributes are indicated with superscripts $()^{layer}_{node, patch}$, like $x^l_{i,p}$ is the node feature of node $i$ in patch $p$ at layer $l$. 
\vspace{5pt}

\noindent\textbf{Definition}
\label{def:lbo}
The volumetric Laplace-Beltrami Operator (LBO), $\Delta_{tet}$, on a tetrahedral mesh $M$ with vertices $\{v_i\}_{i=1}^n$ satisfies the eigen-system $\Delta_{tet} f = -\lambda f$ for a real-valued function $f \in C^2$. Its discrete approximation~\cite{wang2017towards} is:
\begin{equation}
    \Delta f(v_i) = \frac{1}{d_i} \sum_{j \in N(i)} k_{i,j} (f(v_i) - f(v_j)),
\end{equation}
where $N(i)$ are vertices adjacent to $v_i$, $d_i$ is the volume of adjacent tetrahedra, and $k_{i,j}$ depends on edge lengths and dihedral angles. The stiffness matrix $A = W - K$, with $W = \text{diag}(w_1, \dots, w_n)$ and $w_i = \sum_{j \in N(i)} k_{i,j}$. The lumped discrete LBO is $L_{tet} = D^{-1}A$ where $D = \text{diag}(d_1, \dots, d_n)$.

\begin{figure}[t]
\includegraphics[width=\textwidth]{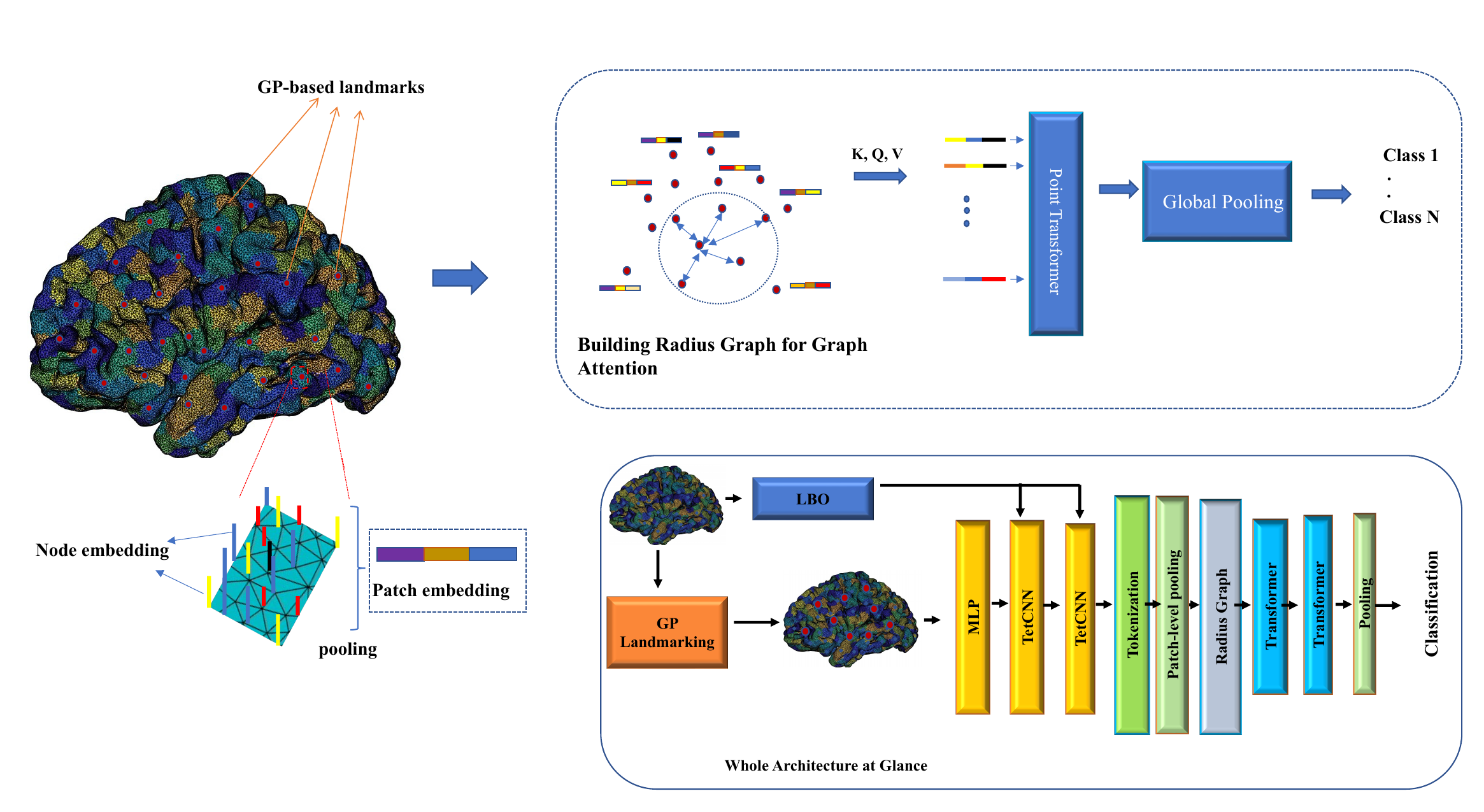}
\caption{The architecture of our model begins with identifying patch centers using Gaussian Process (GP)-based landmarks. Local-aware features are then learned through TetCNN, followed by constructing a radius graph to compute sparse attention over the previously learned tokens. Additionally, we pre-compute the Laplace-Beltrami Operator (LBO) and landmarks to streamline the process.} \label{fig:architecture}
\end{figure}

\subsection{Gaussian Process-based Anatomical Landmarking}

Anatomical landmarking was initially developed for dimensionality reduction, or saliency detection~\cite{liu2016shape}. A recent study proposed a Gaussian process-based method for automatically generating anatomical landmarks~\cite{fan2020convolutional}. To be specific, a multi-frequency multi-phase periodic diffusion kernel (mmPDK) was designed to encode high-quality geometric features into the prior knowledge of a Gaussian process (GP) model, resulting in a geometry-aware landmark generation framework for generating an arbitrary number of landmarks. In this study, we leveraged the proposed landmark generation algorithm to predefine a group of landmarks for each tetrahedral mesh. These generated landmarks were then incorporated into our GNN-based model as additional inputs. The overall architecture of the model is also illustrated in Fig ~\ref{fig:architecture}.

\begin{figure}[t]
\includegraphics[width=\textwidth]{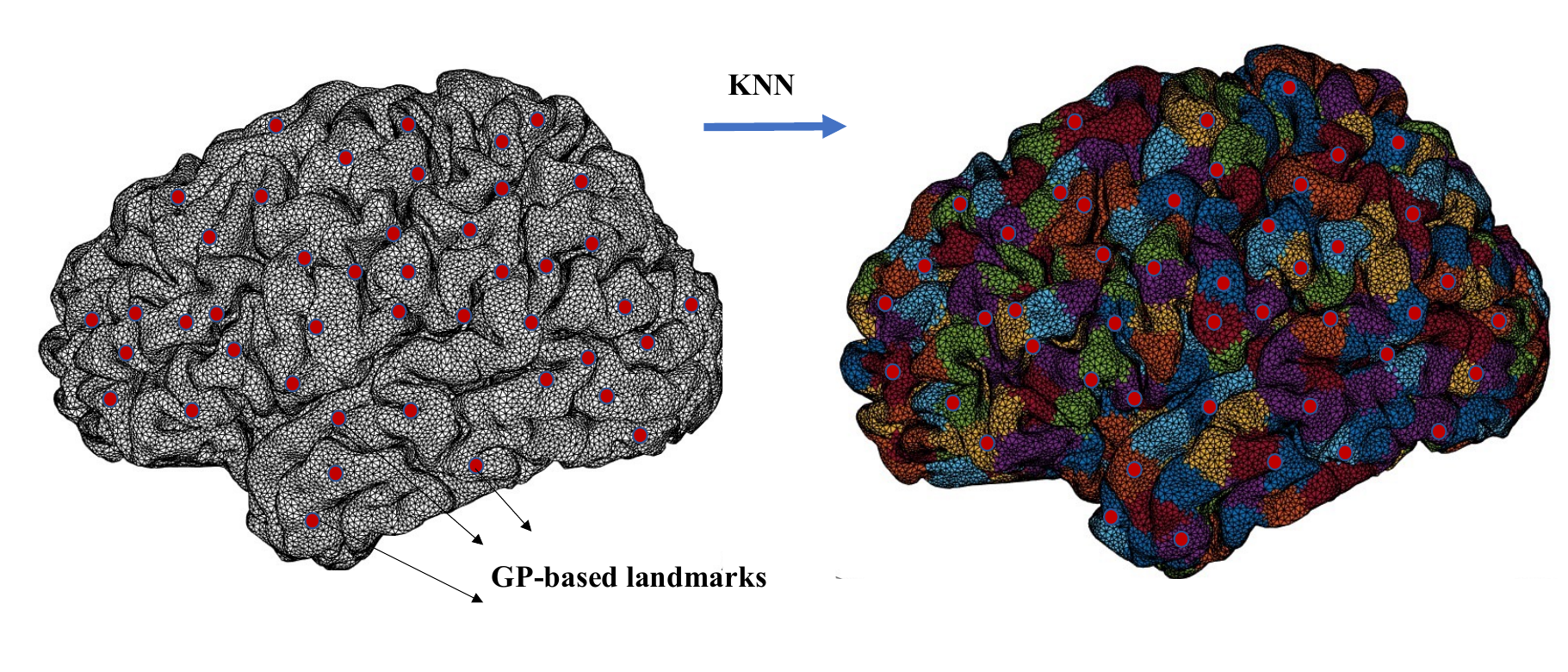}
\caption{Visualization of tokenization of the tetrahedral mesh. We assigned each node of the mesh the the nearest super node to create our tokens, resulting in patchwise visualization of the mesh. The supernodes are located in the centers of each patch. Each color represent unique patch (token).} \label{fig2}
\end{figure}

\subsection{Tokenization}

Based on our landmarks (super nodes), we use the K-nearest Neighbor algorithm to assign each node of the mesh to these super nodes to create our tokens. We assign the labels to each node based on their proximity to their immediate close super nodes. Unlike in images with a uniform grid structure, here on volumetric meshes, patches generated in any fashion always have different sizes in terms of nodes. Figure ~\ref{fig2} better illustrates the way how tokens are generated. After creating tokens, the center of each patch is also saved for transformer processing.

\subsection{Node and Patch-Level Embeddings}
Unlike traditional ViT-style transformers that rely on fixed-size patches with a consistent number of nodes per patch, we propose leveraging patch embeddings without relying on padding, accommodating inputs with highly variable node counts. For patch embedding, we first propose learning primary local feature representation using TetCNN layers~\cite{farazi2023tetcnn}. TetCNN is tailored for graph convolution on tetrahedral meshes by substituting the graph Laplacian operator with a customized volumetric Laplacian Beltrami Operator (volumetric LBO). The computational overhead of the graph convolution is substantially reduced by approximating the operation with spectral filtering of mesh signals, accomplished by Chebyshev polynomial approximation~\cite{farazi2023tetcnn} This helps to learn further features not limited to raw 3D coordinates. In order to represent one patch, we use pooling to average the node features of that patch. We can define node-level embeddings and patch-level embeddings based on the following:

\begin{equation}\label{eq_tet}
   x_{i,0} = f_{\theta}([ x_{in}])
\end{equation}
\begin{equation}\label{eq_tet2}
   x_{i,\ell+1} = \sum_{m=0}^{K}\theta_{m}T_{m}(L_{tet})x_{i,\ell}, 
\end{equation}
For the first layer, we employ a point-wise MLP to project the raw node features—specifically the 3D coordinates—into a higher-dimensional space using an MLP layer followed by an activation function, i.e. $f_{\theta}$. Here, $K$ defines the maximum polynomial order, $\theta_m$ are learnable weights for each order, and $T_m(L_{tet})$ represents the $m$-th Chebyshev polynomial of the tetrahedral Laplacian $L_{tet}$.

Next, we use a simple pooling strategy per patch to generate each token embedding with uniform embedding size across all $|P|$ patches.

\begin{equation}
    X^p = \frac{1}{|p|}\sum_{i \in p} { x_{i}}
\end{equation}
In this equation $p$ is an arbitrary patch on the mesh, and $|p|$ shows the number of nodes in the patch. $X$ shows the patch embedding. With these embeddings, the tokens are now prepared. One approach is to add positional encodings and input them into a conventional transformer. However, due to the computational cost of full self-attention and the opportunity to leverage geometric information, a graph-based approach for calculating attention is more suitable.

\subsection{Super Node Representation}

A super node is represented as a point in 3D space equipped with an embedding in a higher-dimensional feature space. Each super node is denoted as $s_i = (p_i, e_i)$, where $p_i \in \mathbb{R}^3$ corresponds to the spatial coordinates $(x, y, z)$ of the super node, and $e_i \in \mathbb{R}^d$ is its feature embedding in a $d$-dimensional space. The set of all super nodes is defined as:
\[
\mathcal{S} = \{s_1, s_2, \ldots, s_N\}, \quad s_i = (p_i, e_i), \quad p_i \in \mathbb{R}^3, \; e_i \in \mathbb{R}^d.
\]

\subsubsection{Radius Graph Construction}
A radius graph \( G = (\mathcal{V}, \mathcal{E}) \) is constructed by defining edges between super nodes that are within a fixed radius \( r \). Specifically, for each super node \( s_i \), an edge \( (i, j) \in \mathcal{E} \) exists if the Euclidean distance between their spatial coordinates \( p_i \) and \( p_j \) satisfies:
\[
\| p_i - p_j \|_2 \leq r,
\]
where \( \| \cdot \|_2 \) denotes the \( \ell_2 \)-norm. The graph is defined as:
\[
G = (\mathcal{V}, \mathcal{E}), \quad \mathcal{E} = \{(i, j) \mid \| p_i - p_j \|_2 \leq r \}.
\]

\subsubsection{Relative Positional Encoding}
For each edge \( (i, j) \), the relative positional encoding~\cite{wu2022point,park2022grpe,zhao2021point} \( \Delta p_{ij} \) is computed as:
\[
\Delta p_{ij} = p_j - p_i,
\]
which captures the spatial relationship between neighboring super nodes. This positional encoding is incorporated into the attention mechanism.

\subsection{Attention Mechanism}

The proposed transformer-based framework leverages an attention mechanism designed for tetrahedral meshes. The attention mechanism integrates both feature similarity and geometric relationships by computing attention scores for each edge $(i, j)$ in the graph. The attention score is given by:
\begin{equation}
    \text{Attention}(q_i, k_j, \Delta p_{ij}) = \sigma \big( \phi(q_i, k_j) + \delta(\Delta p_{ij}) \big),
\end{equation}
where:
\begin{itemize}
    \item $q_i$ and $k_j$ are the query and key vectors of super nodes $i$ and $j$, respectively.
    \item $\phi$: A learnable function capturing feature similarity between $q_i$ and $k_j$.
    \item $\delta$: A learnable function encoding relative positional information $\Delta p_{ij}$.
    \item $\sigma$: A normalization function, such as softmax, ensuring valid attention scores.
\end{itemize}

In this formulation, $\Delta p_{ij}$ encodes geometric and topological relationships between super nodes $i$ and $j$, allowing the model to capture both local and global structure in tetrahedral meshes. Unlike prior works focused on point clouds or fixed-sized patches, our method adapts to varying mesh topologies by incorporating anatomical landmarks as super nodes and constructing task-specific graphs for attention computation.

\subsection{Integrating BBBMs}
After extracting the graph features, we integrated blood-based biomarkers by concatenating the BBBM values with the feature vector in the first fully connected layer, resulting in $d+1$ dimension inputs.

\subsection{Computational Efficiency}
By using a radius graph, the Point Transformer focuses only on local neighborhoods, significantly reducing computational complexity compared to full self-attention over all points. The sparsity of the graph ensures that the attention mechanism is applied efficiently.

\section{Experimental Design}
\subsection{Datasets}
Data used in this study was downloaded the Alzheimer's Disease Neuroimaging Initiative (ADNI) database~\cite{jack2008alzheimer}. We evaluated our model on two subsets. For  A total of 947 samples from 906 individuals were contained in the dataset used for AD classification experiments, including 313 with Alzheimer’s disease (AD), 402 with mild cognitive impairment (MCI), and 229 being cognitively normal (CN). In addition, the dataset for brain amyloid positivity classification contained 909 data samples with matching T1-weighted brain MRI, PET scan and pTau-217 measured by PrecivityAD2 released by C2N diagnostics.  (Table~\ref{tab:1}). All samples within the same dataset were processed using the same protocol.

\begin{table}[t]
    \centering
    \begin{tabular}{c|c|c||c|c|c|c|c|c}
        \hline
        \multicolumn{3}{c||}{AD Classification} & \multicolumn{3}{c|}{A$\beta$+} & \multicolumn{3}{|c}{A$\beta$-} \\
        \hline
        N(AD) & N(MCI) & N(CN) & Low & Medium & High & Low & Medium & High \\
        \hline
        313 & 402 & 229 & 19 & 70 & 265 & 335 & 96 & 23 \\
        \hline
    \end{tabular}
    \vspace{7pt}
    \caption{Dataset used for AD classification task and brain amyloid positivity classification task.  Note that most A$\beta$+ samples have high pTau-217 levels, and most A$\beta$- samples have low pTau-217 levels. We aimed at distinguishing A$\beta$ positivity for the medium risk group besides the overall classification performance.}
    \label{tab:1}
\end{table}

\subsection{Data Preprocessing}
The collected MRI images were first processed by Freesurfer to reconstruct the pial and white surfaces. We then generated the tetrahedral meshes using the TetGen pipeline~\cite{hang2015tetgen}, resulting in tetrahedral meshes containing around 70k-100k vertices. Finally, 1000 anatomical landmarks were extracted from each tetrahedral mesh using a GP-based landmarking generation pipeline. The ground truth class labels were identified by computing the Centiloid values from PET images. Specifically, the PET images were processed by PET Unified Pipeline (PUP), and the positivity threshold was set to $Centiloid > 20$, as proposed and validated in previous studies~\cite{klunk2015centiloid}. In addition, we applied a two-threshold approach to classify individuals into low, medium and high risk of being amyloid positive ~\cite{Figdore:2cutoff2024}. The two thresholds were 1.53 and 2.602, determined by 95\% sensitivity and 95\% specificity, respectively. Since the pTau-217 alone has already achieved significant success as a sole predictor of brain amyloid positivity (AUC=0.94 (95\% CI 0.92–0.97)), but not in the medium group (as shown in our results), we focus our effort on the medium group in our study~\cite{arranz2024diagnostic}.

\subsection{Model Training}
Prior to feeding data into our model, we followed the approach proposed in~\cite{farazi2023tetcnn} to pre-compute the mass matrix and cotangent matrix for each tetrahedral mesh. We initially applied two layers of TetCNN for feature learning. To construct the graph for self-attention computation, we generated a radius graph with $r=0.5$ using the PyG library~\cite{fey2019fast}. Subsequently, we employed two Point Transformer layers~\cite{fey2019fast,zhao2021point} on the generated graph, with a hidden dimension size of $128$ for all layers. Finally, a global pooling operation was applied for both final classification and patch embeddings. The model was trained using binary cross-entropy loss, the ADAM optimizer with a learning rate of $10^{-4}$, weight decay of $10^{-4}$, and a total of $500$ epochs. Due to the small batch size of $2$, gradient accumulation was employed to improve training efficiency.

\subsection{Evaluation Protocol}
For AD classification, we evaluated our model performance on classification tasks between all pairs of diagnosis groups. We compared our LETetCNN model with baselines including ChebyNet~\cite{defferrard2016convolutional} and GAT~\cite{velivckovic2017graph}. For brain amyloid positivity classification, we compared our LETetCNN model and the incorporated LETetCNN + pTau-217 model with three additional baseline models besides the aforementioned GNN-based models, including logistic regression models using hippocampal volume, pTau-217 alone, and hippocampal volume + pTau-217 as predictors. For each classification task, we calculated the accuracy, sensitivity and specificity and compared the metrics among all models.

\section{Results}

\begin{table}[t]
    \centering
    \begin{tabular}{c|c|c|c|c|c|c|c|c|c}
        \hline
        & \multicolumn{3}{|c|}{AD vs CN} & \multicolumn{3}{|c|}{AD vs MCI} & \multicolumn{3}{|c}{MCI vs CN} \\
        \hline
        Model & ACC & SEN & SPE & ACC & SEN & SPE & ACC & SEN & SPE \\
        \hline
        ChebyNet & 0.870 & 0.881 & 0.850 & 0.703 & \textbf{0.790} & 0.616 & 0.735 & 0.778 & 0.667 \\
        \hline
        GAT & 0.858 & 0.873 & 0.836 & 0.727 & 0.630 & 0.773 & 0.722 & 0.763 & 0.660 \\
        \hline
        TetCNN & 0.876 & 0.886 & 0.859 & 0.709 & 0.660 & 0.769 & 0.730 & 0.761 & 0.700 \\
        \hline
        \textbf{LETetCNN} & \textbf{0.917} & \textbf{0.915} & \textbf{0.920} & \textbf{0.755} & 0.615 & \textbf{0.835} & \textbf{0.794} & \textbf{0.778} & \textbf{0.822} \\
        \hline
    \end{tabular}

    \vspace{8pt}
    \caption{Classification results among clinical diagnosis groups. Our LETetCNN model achieved the best performance across all classification tasks. In general, AD vs MCI and MCI vs CN were less separable by all classifiers.}
    \label{tab:diagnosisgroup}
\end{table}

\subsection{Alzheimer’s Disease Classification}
As shown in Table~\ref{tab:diagnosisgroup}, our model consistently outperformed existing methods across all clinical diagnosis group pairs, with varied degrees of improvement observed across groups. Specifically, we achieved 91.7\% accuracy in AD vs CN classification, and 75.5\% accuracy in AD vs MCI classification, and 79.4\% in MCI vs CN classification. While the individuals in AD diagnosis group typically exhibit a higher stage of dementia, leading to a significant geometric distinction between the AD and MCI/CN groups, discriminating between MCI and CN subgroups based on cortical structural change poses more challenge. Therefore, our model’s superior classification performance, particularly in MCI vs CN comparisons, demonstrates the effectiveness and robustness of our method. Moreover, integrating anatomical landmarks further enhanced the model's performance, highlighting the strength of our approach.

\begin{table}[t]
    \centering
    \begin{tabular}{c|c|c|c}
        \hline
        Model & ACC & SEN & SPE \\
        \hline\hline
        Hippocampal Volume & 0.450 & 0.529 & 0.391 \\
        \hline
        pTau-217 & 0.675 & 0.750 & 0.625 \\
        \hline
        Hippocampal Volume + pTau-217 & 0.675 & 0.750 & 0.625 \\
        \hline
        ChebyNet & 0.677 & 0.563 & 0.800 \\
        \hline
        GAT & 0.677 & 0.611 & 0.769 \\
        \hline
        TetCNN & 0.690 & 0.684 & 0.694  \\
        \hline
        LETetCNN & 0.758 & 0.785 & 0.733  \\
        \hline
        \textbf{LETetCNN+pTau-217} & \textbf{0.798} & \textbf{0.785} & \textbf{0.811}
    \end{tabular}
    \caption{Amyloid positivity prediction results in the medium risk groups. An integration of LETetCNN and pTau-217 showed enhanced accuracy, outperforming all other baseline classifiers.}
    \label{tab:preAD}
\end{table}

\subsection{Brain Amyloid Positivity Classification}
We generated a total of 909 brain cortical tetrahedral meshes from structural MR images. All samples were from ADNI dataset with matching PET scans and pTau-217 measures. We followed the widely used protocol~\cite{klunk2015centiloid} to determine brain amyloid positivity by calculating Centiloid value for each PET image, and labeled any PET image with $Centiloid > 20$ as amyloid positive. We employed the aforementioned two-threshold method to classify samples into low, medium and high groups, and primarily focus on the medium group, where pTau-217 alone has demonstrated limited classification power. The subset of the medium group contained 70 positive and 96 negative samples. Hippocampal volume was also included as a baseline classifier, given its well-documented correlation with AD-related dementia. However, as shown in Table~\ref{tab:preAD}, hippocampal volume yielded results close to random classification. In comparison, pTau-217 alone achieved 0.675 classification accuracy, with similar performance observed from other GNN-based methods. In contrast, our LETetCNN model outperformaned all baseline models by achieving an accuracy of 0.758. Notably, the performance was further enhanced after integrating pTau-217, reaching a classification accuracy of 0.798 with a balanced prediction (SEN=0.785, SPE=0.811). While further validations are needed, our current results suggest that our work could complement BBBM research, enhancing its accuracy through a cost-effective and less invasive approach.

\subsection{Ablation Studies}
We first investigated the impact of learned features extracted by a few TetCNN layers before they were passed to the transformer. To assess this, we removed these layers and directly feed the raw features into the transformer. In the second ablation stude, we evaluated whether integrating $p$Tau-217 with LETetCNN significantly enhanced performance. This analysis was crucial, as $p$Tau-217 itself was a highly predictive biomarker, and it was important to determine whether the features learned by our network provided additional complementary information. The results of both studies were summarized in Table~\ref{tab:4}. It demonstrated TetCNN learned important geometric features and the proposed framework further enhanced the performance of BBBMs for preclinical AD research.

\begin{table}[t]
    \centering
    \begin{tabular}{c|c|c|c}
        \hline
        Model & ACC & SEN & SPE \\
        \hline\hline
        LE & 0.725 & 0.674 & 0.750 \\
        \hline
        LETetCNN & 0.765 & 0.738 & 0.790  \\        
        \hline
        pTau-217 & 0.675 & 0.750 & 0.625\\ \hline
        \textbf{LETetCNN+pTau-217} & \textbf{0.798} & \textbf{0.785} & \textbf{0.811} \\
        \hline

        \hline
    \end{tabular}
    \vspace{7pt}
    \caption{The ablation study results for amyloid positivity prediction in the medium-risk group indicate that the transformer-only architecture (LE) did not yield competitive performance compared to using geometric features learned from TetCNN as input. Additionally, our proposed approach provided complementary information to enhance the performance of BBBMs.}
    \label{tab:4}
\end{table}

\subsection{Grad-CAM Visualization}
\begin{figure}[t]
\includegraphics[width=\textwidth]{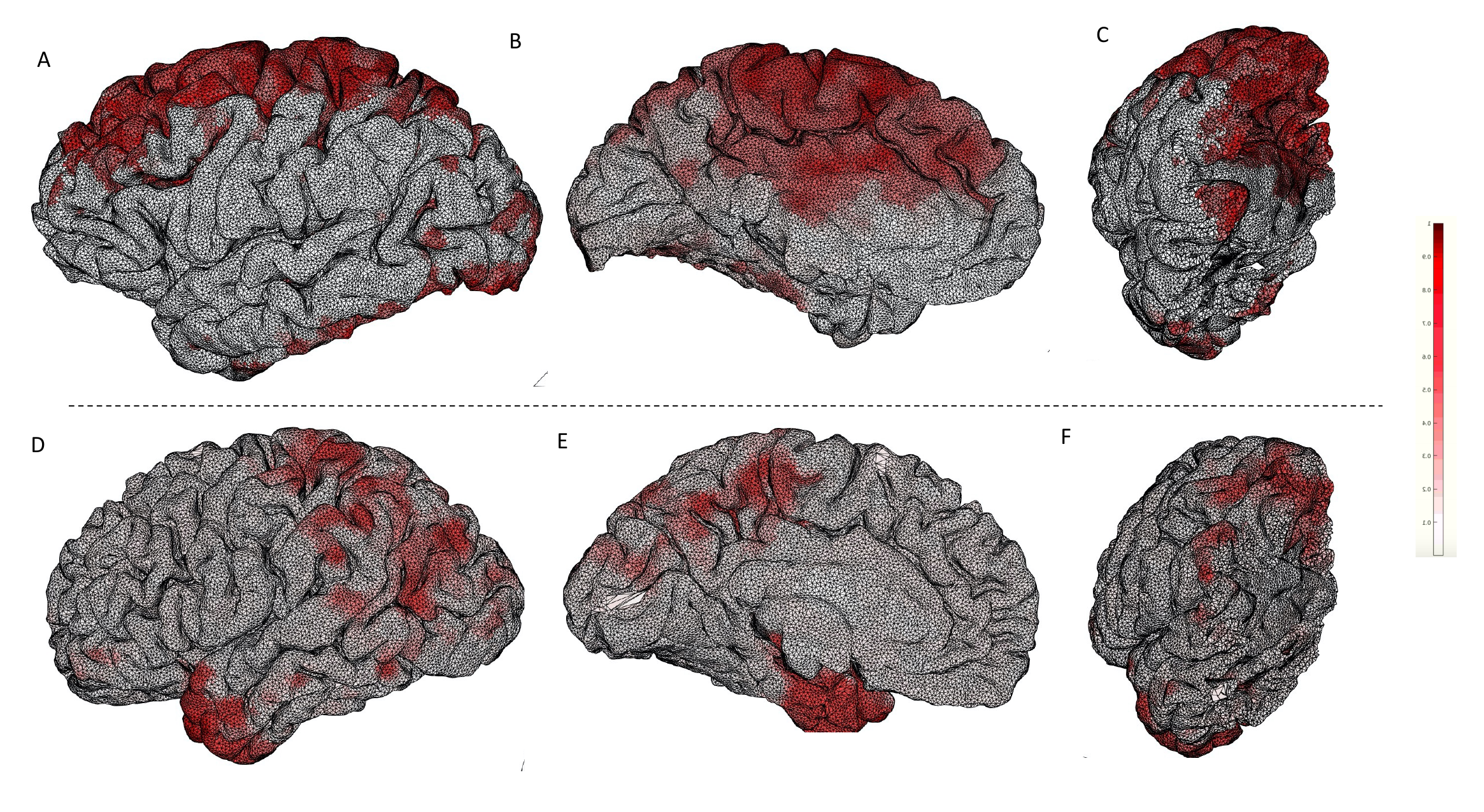}
\caption{Grad-CAM visualization highlighting regions of interest that align with Alzheimer's Disease (AD) pathology. Both results from AD classification task (A-C) and brain amyloid positivity prediction task (D-F) exhibited strong activations in the temporal gyrus and parts of the inferior temporal lobe. Mild activation was also observed in superior frontal cortex, which was related to later-stage disease progression. These results closely align with neuroimaging studies identifying key biomarkers for AD pathology.} \label{fig:3}
\end{figure}

We visualized the activation map of our model prediction for the AD classification task (Fig ~\ref{fig:3}). The highlighted areas were spreading around posterior cingulate and medial temporal lobe, with some less significant regions located in frontal lobe. Medial temporal lobe has been widely reported as the first region affected by AD, and posterior cingulate was also shown to be functionally important in impairment of learning and memory~\cite{jack1998rate,minoshima1997metabolic}. In addition, frontal lobe was also shown to be correlated in late stage AD~\cite{johnson1999clinical}. Therefore, the Grad-CAM heatmap demonstrated high correspondence of pathologically significant regions identified by our model with those from existing literature.

\section{Conclusion and Future Work}

In this work, we proposed a novel transformer-based framework tailored specifically for tetrahedral meshes with varying sizes and topologies. To the best of our knowledge, this is the first approach to extend transformer architectures to tetrahedral meshes, leveraging relative positional encodings and task-specific graph tokenization for efficient and geometry-aware representation learning. Unlike prior methods that rely on fixed-size re-meshing or hierarchical pooling, our approach dynamically adapts to the inherent variability of tetrahedral mesh structures. We evaluated our model on two distinct tasks: clinical AD classification and brain amyloid positivity prediction. We compared our model with baseline classifiers, including ChebyNet and GAT in AD classification, and hippocampal volume and pTau-217 in amyloid positivity prediction. In both experiments, our model outperformed all baseline methods, indicating both effectiveness and generalizability. In particular, our LETetCNN model can be empowered by leveraging the pTau-217 information. With the rapid advancement of BBBM studies in AD research~\cite{therriault2023equivalence,groot2022diagnostic,palmqvist2020discriminative}, our work indicates that traditional brain imaging research will still be useful and relevant.

Compared to TetCNN, which utilizes localized filters defined on the volumetric LBO our method introduces several key advantages. First, the attention mechanism in our framework enables dynamic modeling of global relationships across the mesh, overcoming the locality limitation of TetCNN. While TetCNN layers effectively capture local features, their hierarchical pooling may result in loss of critical geometric details, particularly for meshes with irregular or anisotropic topologies. In contrast, our method retains finer global and local information through attention mechanisms that explicitly consider both feature similarity and geometric relationships.

Second, the incorporation of relative positional encodings allows our framework to encode spatial relationships between nodes and patches directly in the feature space. This improves the model's ability to represent both local geometric features and global structural context. Additionally, our use of anatomical landmarks as super nodes ensures robust and semantically meaningful tokenization, which is particularly beneficial for handling large volumetric datasets.

Empirical results validate the effectiveness of our approach. Our framework achieves superior performance in clinical group classification, with an average of 5\% increase in accuracy compared to TetCNN. Visualizations of attention maps further highlight the ability of our model to focus on critical regions, demonstrating richer feature representations. Furthermore, ablation studies show the significant contributions of relative positional encodings and landmark-based tokenization to overall performance.

In addition to its representational advantages, our framework maintains computational efficiency by leveraging sparse graph structures and localized attention. This scalability makes it well-suited for large-scale tetrahedral meshes, addressing a key limitation of TetCNN and other traditional methods. 


\bibliographystyle{splncs04}
\bibliography{refs}

@Article{Figdore:2cutoff2024,
   Author="Figdore, D. J.  and Griswold, M.  and Bornhorst, J. A.  and Graff-Radford, J.  and Ramanan, V. K.  and Vemuri, P.  and Lowe, V. J.  and Knopman, D. S.  and Jack, C. R.  and Petersen, R. C.  and Algeciras-Schimnich, A. ",
   Title="{{O}ptimizing cutpoints for clinical interpretation of brain amyloid status using plasma p-tau217 immunoassays}",
   Journal="Alzheimers Dement",
   Year="2024",
   Volume="20",
   Number="9",
   Pages="6506--6516",
   Month="Sep"
}

@article{jack2008alzheimer,
  title={The {A}lzheimer's disease neuroimaging initiative {(ADNI)}: {MRI} methods},
  author={Jack Jr, Clifford R and Bernstein, Matt A and Fox, Nick C and Thompson, Paul and Alexander, Gene and Harvey, Danielle and Borowski, Bret and Britson, Paula J and L. Whitwell, Jennifer and Ward, Chadwick and others},
  journal={J Magn Reson Imaging},
  volume={27},
  number={4},
  pages={685--691},
  year={2008},
  publisher={Wiley Online Library}
}

@article{qiu2009epidemiology,
  title={Epidemiology of Alzheimer's disease: occurrence, determinants, and strategies toward intervention},
  author={Qiu, Chengxuan and Kivipelto, Miia and Von Strauss, Eva},
  journal={Dialogues in clinical neuroscience},
  volume={11},
  number={2},
  pages={111--128},
  year={2009},
  publisher={Taylor \& Francis}
}

@article{el2019tip,
  title={Tip of the iceberg: assessing the global socioeconomic costs of Alzheimer’s disease and related dementias and strategic implications for stakeholders},
  author={El-Hayek, Youssef H and Wiley, Ryan E and Khoury, Charles P and Daya, Ritesh P and Ballard, Clive and Evans, Alison R and Karran, Michael and Molinuevo, Jos{\'e} Luis and Norton, Matthew and Atri, Alireza},
  journal={Journal of Alzheimer's Disease},
  volume={70},
  number={2},
  pages={323--341},
  year={2019},
  publisher={IOS Press}
}

@inproceedings{atri2019current,
  title={Current and future treatments in Alzheimer's disease},
  author={Atri, Alireza},
  booktitle={Seminars in neurology},
  volume={39},
  number={02},
  pages={227--240},
  year={2019},
  organization={Thieme Medical Publishers}
}

@article{liss2021practical,
  title={Practical recommendations for timely, accurate diagnosis of symptomatic Alzheimer’s disease (MCI and dementia) in primary care: a review and synthesis},
  author={Liss, JL and Seleri Assun{\c{c}}{\~a}o, S and Cummings, Jeffrey and Atri, A and Geldmacher, DS and Candela, SF and Devanand, DP and Fillit, HM and Susman, J and Mintzer, J and others},
  journal={Journal of internal medicine},
  volume={290},
  number={2},
  pages={310--334},
  year={2021},
  publisher={Wiley Online Library}
}

@article{dickerson2009cortical,
  title={The cortical signature of Alzheimer's disease: regionally specific cortical thinning relates to symptom severity in very mild to mild AD dementia and is detectable in asymptomatic amyloid-positive individuals},
  author={Dickerson, Bradford C and Bakkour, Akram and Salat, David H and Feczko, Eric and Pacheco, Jenni and Greve, Douglas N and Grodstein, Fran and Wright, Christopher I and Blacker, Deborah and Rosas, H Diana and others},
  journal={Cerebral cortex},
  volume={19},
  number={3},
  pages={497--510},
  year={2009},
  publisher={Oxford University Press}
}

@article{atri2019alzheimer,
  title={The Alzheimer’s disease clinical spectrum: diagnosis and management},
  author={Atri, Alireza},
  journal={Medical Clinics},
  volume={103},
  number={2},
  pages={263--293},
  year={2019},
  publisher={Elsevier}
}

@article{fawaz2021benchmarking,
  title={Benchmarking geometric deep learning for cortical segmentation and neurodevelopmental phenotype prediction},
  author={Fawaz, Abdulah and Williams, Logan ZJ and Alansary, Amir and Bass, Cher and Gopinath, Karthik and da Silva, Mariana and Dahan, Simon and Adamson, Chris and Alexander, Bonnie and Thompson, Deanne and others},
  journal={bioRxiv},
  pages={2021--12},
  year={2021},
  publisher={Cold Spring Harbor Laboratory}
}

@article{hanocka2019meshcnn,
  title={Meshcnn: a network with an edge},
  author={Hanocka, Rana and Hertz, Amir and Fish, Noa and Giryes, Raja and Fleishman, Shachar and Cohen-Or, Daniel},
  journal={ACM Transactions on Graphics (ToG)},
  volume={38},
  number={4},
  pages={1--12},
  year={2019},
  publisher={ACM New York, NY, USA}
}

@article{chentanez2007liquid,
  title={Liquid simulation on lattice-based tetrahedral meshes},
  author={Chentanez, Nuttapong and Feldman, Bryan E and Labelle, Fran{\c{c}}ois and O’Brien, James F and Shewchuk, Jonathan R},
  year={2007}
}

@inproceedings{farazi2023tetcnn,
  title={{TetCNN}: Convolutional neural networks on tetrahedral meshes},
  author={Farazi, Mohammad and Yang, Zhangsihao and Zhu, Wenhui and Qiu, Peijie and Wang, Yalin},
  booktitle={International Conference on Information Processing in Medical Imaging},
  pages={303--315},
  year={2023},
  organization={Springer}
}

@inproceedings{wang2004volumetric,
  title={Volumetric harmonic brain mapping},
  author={Wang, Yalin and Gu, Xianfeng and Chan, Tony F and Thompson, Paul M and Yau, Shing-Tung},
  booktitle={2004 2nd IEEE International Symposium on Biomedical Imaging: Nano to Macro (IEEE Cat No. 04EX821)},
  pages={1275--1278},
  year={2004},
  organization={IEEE}
}

@article{wang2017towards,
  title={Towards a holistic cortical thickness descriptor: heat kernel-based grey matter morphology signatures},
  author={Wang, Gang and Wang, Yalin and Alzheimer's Disease Neuroimaging Initiative and others},
  journal={Neuroimage},
  volume={147},
  pages={360--380},
  year={2017},
  publisher={Elsevier}
}

@article{velivckovic2017graph,
  title={Graph attention networks},
  author={Veli{\v{c}}kovi{\'c}, Petar and Cucurull, Guillem and Casanova, Arantxa and Romero, Adriana and Lio, Pietro and Bengio, Yoshua},
  journal={arXiv preprint arXiv:1710.10903},
  year={2017}
}

@inproceedings{fan2020convolutional,
  title={Convolutional bayesian models for anatomical landmarking on multi-dimensional shapes},
  author={Fan, Yonghui and Wang, Yalin},
  booktitle={Medical Image Computing and Computer Assisted Intervention--MICCAI 2020: 23rd International Conference, Lima, Peru, October 4--8, 2020, Proceedings, Part IV 23},
  pages={786--796},
  year={2020},
  organization={Springer}
}

@article{liu2016shape,
  title={Shape context based mesh saliency detection and its applications: A survey},
  author={Liu, Xianyong and Liu, Ligang and Song, Weijie and Liu, Yanping and Ma, Lizhuang},
  journal={Computers \& Graphics},
  volume={57},
  pages={12--30},
  year={2016},
  publisher={Elsevier}
}

@article{defferrard2016convolutional,
  title={Convolutional neural networks on graphs with fast localized spectral filtering},
  author={Defferrard, Micha{\"e}l and Bresson, Xavier and Vandergheynst, Pierre},
  journal={Advances in neural information processing systems},
  volume={29},
  year={2016}
}

@article{hang2015tetgen,
  title={TetGen, a Delaunay-based quality tetrahedral mesh generator},
  author={Hang, Si},
  journal={ACM Trans. Math. Softw},
  volume={41},
  number={2},
  pages={11},
  year={2015}
}

@article{nakamura2018high,
  title={High performance plasma amyloid-$\beta$ biomarkers for Alzheimer’s disease},
  author={Nakamura, Akinori and Kaneko, Naoki and Villemagne, Victor L and Kato, Takashi and Doecke, James and Dor{\'e}, Vincent and Fowler, Chris and Li, Qiao-Xin and Martins, Ralph and Rowe, Christopher and others},
  journal={Nature},
  volume={554},
  number={7691},
  pages={249--254},
  year={2018},
  publisher={Nature Publishing Group UK London}
}

@article{ashton2023plasma,
  title={Plasma and CSF biomarkers in a memory clinic: head-to-head comparison of phosphorylated tau immunoassays},
  author={Ashton, Nicholas J and Puig-Pijoan, Albert and Mil{\`a}-Alom{\`a}, Marta and Fern{\'a}ndez-Lebrero, Aida and Garc{\'\i}a-Escobar, Greta and Gonz{\'a}lez-Ortiz, Fern{\'a}ndo and Kac, Przemys{\l}aw R and Brum, Wagner S and Benedet, Andr{\'e}a L and Lantero-Rodriguez, Juan and others},
  journal={Alzheimer's \& Dementia},
  volume={19},
  number={5},
  pages={1913--1924},
  year={2023},
  publisher={Wiley Online Library}
}

@article{therriault2023equivalence,
  title={Equivalence of plasma p-tau217 with cerebrospinal fluid in the diagnosis of Alzheimer's disease},
  author={Therriault, Joseph and Servaes, Stijn and Tissot, C{\'e}cile and Rahmouni, Nesrine and Ashton, Nicholas J and Benedet, Andr{\'e}a Lessa and Karikari, Thomas K and Macedo, Arthur C and Lussier, Firoza Z and Stevenson, Jenna and others},
  journal={Alzheimer's \& Dementia},
  volume={19},
  number={11},
  pages={4967--4977},
  year={2023},
  publisher={Wiley Online Library}
}

@article{groot2022diagnostic,
  title={Diagnostic and prognostic performance to detect Alzheimer’s disease and clinical progression of a novel assay for plasma p-tau217},
  author={Groot, Colin and Cicognola, Claudia and Bali, Divya and Triana-Baltzer, Gallen and Dage, Jeffrey L and Pontecorvo, Michael J and Kolb, Hartmuth C and Ossenkoppele, Rik and Janelidze, Shorena and Hansson, Oskar},
  journal={Alzheimer's research \& therapy},
  volume={14},
  number={1},
  pages={67},
  year={2022},
  publisher={Springer}
}

@article{palmqvist2020discriminative,
  title={Discriminative accuracy of plasma phospho-tau217 for Alzheimer disease vs other neurodegenerative disorders},
  author={Palmqvist, Sebastian and Janelidze, Shorena and Quiroz, Yakeel T and Zetterberg, Henrik and Lopera, Francisco and Stomrud, Erik and Su, YI and Chen, Yinghua and Serrano, Geidy E and Leuzy, Antoine and others},
  journal={Jama},
  volume={324},
  number={8},
  pages={772--781},
  year={2020},
  publisher={American Medical Association}
}

@article{klunk2015centiloid,
  title={The Centiloid Project: standardizing quantitative amyloid plaque estimation by PET},
  author={Klunk, William E and Koeppe, Robert A and Price, Julie C and Benzinger, Tammie L and Devous Sr, Michael D and Jagust, William J and Johnson, Keith A and Mathis, Chester A and Minhas, Davneet and Pontecorvo, Michael J and others},
  journal={Alzheimer's \& dementia},
  volume={11},
  number={1},
  pages={1--15},
  year={2015},
  publisher={Elsevier}
}

@article{arranz2024diagnostic,
  title={Diagnostic performance of plasma pTau217, pTau181, A$\beta$1-42 and A$\beta$1-40 in the LUMIPULSE automated platform for the detection of Alzheimer disease},
  author={Arranz, Javier and Zhu, Nuole and Rubio-Guerra, Sara and Rodr{\'\i}guez-Baz, {\'I}{\~n}igo and Ferrer, Rosa and Carmona-Iragui, Mar{\'\i}a and Barroeta, Isabel and Ill{\'a}n-Gala, Ignacio and Santos-Santos, Miguel and Fortea, Juan and others},
  journal={Alzheimer's Research \& Therapy},
  volume={16},
  number={1},
  pages={139},
  year={2024},
  publisher={Springer}
}

@inproceedings{dosovitskiy2020image,
  title={An Image is Worth 16x16 Words: Transformers for Image Recognition at Scale},
  author={Dosovitskiy, Alexey and Beyer, Lucas and Kolesnikov, Alexander and Weissenborn, Dirk and Zhai, Xiaohua and Unterthiner, Thomas and Dehghani, Mostafa and Minderer, Matthias and Heigold, Georg and Gelly, Sylvain and others},
  booktitle={Proceedings of the International Conference on Machine Learning (ICML)},
  year={2021}
}

@inproceedings{zhao2021point,
  title={Point Transformer},
  author={Zhao, Hengshuang and Huang, Li and Gong, Yuan and Wang, Changhu and Lin, Wenchao and Shin, Ren and Snavely, Noah and Shi, Jianping},
  booktitle={Proceedings of the IEEE/CVF Conference on Computer Vision and Pattern Recognition (CVPR)},
  year={2021}
}

@inproceedings{vaswani2017attention,
  title={Attention is All You Need},
  author={Vaswani, Ashish and Shazeer, Noam and Parmar, Niki and Uszkoreit, Jakob and Jones, Llion and Gomez, Aidan N and Kaiser, {\L}ukasz and Polosukhin, Illia},
  booktitle={Advances in Neural Information Processing Systems (NeurIPS)},
  year={2017}
}

@article{wu2022point,
  title={Point transformer v2: Grouped vector attention and partition-based pooling},
  author={Wu, Xiaoyang and Lao, Yixing and Jiang, Li and Liu, Xihui and Zhao, Hengshuang},
  journal={Advances in Neural Information Processing Systems},
  volume={35},
  pages={33330--33342},
  year={2022}
}

@article{fey2019fast,
  title={Fast graph representation learning with PyTorch Geometric},
  author={Fey, Matthias and Lenssen, Jan Eric},
  journal={arXiv preprint arXiv:1903.02428},
  year={2019}
}

@article{park2022grpe,
  title={Grpe: Relative positional encoding for graph transformer},
  author={Park, Wonpyo and Chang, Woonggi and Lee, Donggeon and Kim, Juntae and Hwang, Seung-won},
  journal={arXiv preprint arXiv:2201.12787},
  year={2022}
}

@article{minoshima1997metabolic,
  title={Metabolic reduction in the posterior cingulate cortex in very early Alzheimer's disease},
  author={Minoshima, Saroshi and Giordani, Bruno and Berent, Stanley and Frey, Kirk A and Foster, Norman L and Kuhl, David E},
  journal={Annals of Neurology: Official Journal of the American Neurological Association and the Child Neurology Society},
  volume={42},
  number={1},
  pages={85--94},
  year={1997},
  publisher={Wiley Online Library}
}

@article{jack1998rate,
  title={Rate of medial temporal lobe atrophy in typical aging and Alzheimer's disease},
  author={Jack Jr, Clifford R and Petersen, Ronald C and Xu, Yuecheng and O'Brien, Peter C and Smith, Glenn E and Ivnik, Robert J and Tangalos, Eric G and Kokmen, Emre},
  journal={Neurology},
  volume={51},
  number={4},
  pages={993--999},
  year={1998},
  publisher={AAN Enterprises}
}

@article{johnson1999clinical,
  title={Clinical and pathological evidence for a frontal variant of Alzheimer disease},
  author={Johnson, Julene K and Head, Elizabeth and Kim, Ronald and Starr, Arnold and Cotman, Carl W},
  journal={Archives of neurology},
  volume={56},
  number={10},
  pages={1233--1239},
  year={1999},
  publisher={American Medical Association}
}

\end{document}